\documentclass[twocolumn,A4paper,showpacs,preprintnumbers,amsmath,amssymb]{revtex4-1}
\usepackage{graphicx}
\usepackage{dcolumn}
\usepackage{bm}

\begin{document}

\title{Effect of transport-induced charge inhomogeneity on point-contact Andreev reflection spectra at ferromagnet-superconductor interfaces}

\author{Ya-Fen Hsu$^1$, Tian-Wei Chiang$^{1,2}$, Guang-Yu Guo$^{3,1,}$\footnote{E-mail address: gyguo@phys.ntu.edu.tw}, 
Shang-Fan Lee$^{2,}$\footnote{E-mail address: leesf@phys.sinica.edu.tw}, Jun-Jih Liang$^4$ 
}

\affiliation{
$^1$Department of Physics and Center for Theoretical Sciences, National Taiwan University, Taipei 10617, Taiwan\\
$^2$Institute of Physics, Academia Sinica, Taipei 11529, Taiwan\\
$^3$Graduate Institute of Applied Physics, National Chengchi University, Taipei 11605, Taiwan\\
$^4$Department of Physics, Fu Jen Catholic University, Taipei 24205, Taiwan
}

\begin{abstract}
We investigate the transport properties of a ferromagnet-superconductor interface
within the framework of a modified three-dimensional Blonder-Tinkham-Klapwijk formalism.
In particular, we propose that charge inhomogeneity forms via two unique transport mechanisms, namely,
evanescent Andreev reflection and evanescent quasiparticle transmission.
Furthermore, we take into account the influence of charge inhomogeneity on the interfacial barrier potential
and calculate the conductance as a function of bias voltage.
Point-contact Andreev reflection (PCAR) spectra often show dip structures, large zero-bias conductance enhancement,
and additional zero-bias conductance peak.
Our results indicate that transport-induced charge inhomogeneity could be
a source of all these anomalous characteristics of the PCAR spectra. 

\end{abstract}

\maketitle

\section{Introduction}

Measurement of spin polarization is an important subject in spintronics.
The point-contact Andreev reflection (PCAR) technique
has been widely used as a simple and powerful method in the determination of
conduction electron spin polarization in a broad range of ferromagnets\cite{Upadhyay,Soulen,Ji01}.
The principle behind the spin polarization measurement by the PCAR spectroscopy is based on
the fact that the Andreev reflection\cite{Andreev} probability at a ferromagnet-superconductor interface
is limited by the minority spin carrier density at the Fermi level in the ferromagnet.
In the limit of a clean ballistic ferromagnet-superconductor contact, the spin polarization
$P$ is simply determined by the ratio $G(V=0)/G_N$ of the conductances $G(V=0)$ and $G_N$
at zero and high bias voltage, respectively, which was obtained by decomposing the current into a fully polarized
part and a fully unpolarized part.\cite{Upadhyay,Soulen}
However, for real ferromagnet-superconductor contacts, which are generally not in the clean limit,
accurate determination of $P$ is nontrivial and usually requires a careful analysis of the complete
conductance-voltage ($G-V$) curve.\cite{Ji01}
Blonder {\it et al.} \cite{Blonder} developed a theory (known as the BTK model)
that takes into account the Andreev reflection, for the experimental $G-V$ curves of normal
metal-superconductor contacts with different behaviors ranging from metallic to tunnel junction.
Later, Strijkers {\it et al.} \cite{Strijkers} extended the BTK model to a ferromagnet-superconductor
contact by including the effects of the spin polarization in the ferromagnet.
The extended BTK model of Strijkers {\it et al.} has been applied to analyze the $G-V$ curves of
many ferromagnet-superconductor contacts and also to determine the spin polarization in the ferromagnets
in the contacts.\cite{Ji01,Strijkers}
In addition to the original one-dimensional (1-D) BTK model\cite{Blonder} and the extended 1-D
BTK model\cite{Strijkers},
several extended three-dimensional (3-D) BTK models for both ballistic and diffusive regimes
have also been proposed.\cite{Golubov,Chalsani,Mortensen,Mazin}

Interestingly, anomalous phenomena such as the dip structures
\cite{Soulen,Nowack,Srikanth,DeWilde1,DeWilde2,Walti,Laube,Mao,Gourgy,Daghero},
the zero-bias conductance peak (ZBCP) \cite{Srikanth,Kastalsky,Magnee,Quirion},
and the large zero-bias enhancement \cite{Mao}, may appear in the PCAR spectra
of a wide variety of point contact systems.
For example, Fig. 1(a) shows that unexpected dips emerge when the bias voltage $V$ approaches the superconducting energy gap $\Delta$.
In addition, a narrow peak occurs at $V=0$ (known as the ZBCP), as shown in Fig. 1(b) and (c).
The large zero-bias enhancement may result in the normalized zero-bias conductance being larger than 2.0 [see Fig. 1(a)].
These characteristics cannot be explained by the above-mentioned extended BTK models
\cite{Blonder,Strijkers,Soulen,Golubov,Chalsani,Mortensen,Mazin}
and thus other effects must be taken into account.

\begin{figure}
\includegraphics[width=8cm]{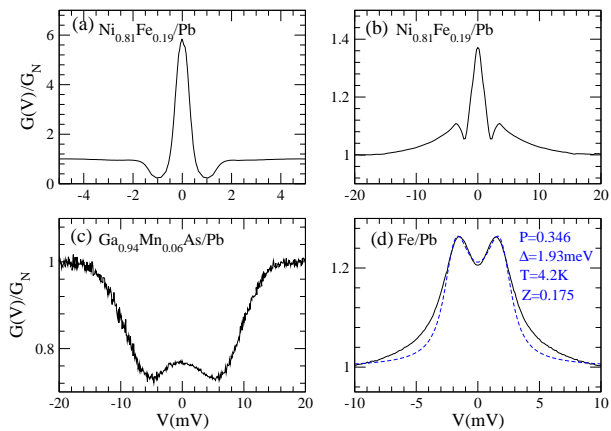}
\caption{(Color online) The anomalous PCAR spectra
from our own measurements at $T=4.2$K (solid lines) (a-c). 
(a) Dip structures for a Ni$_{0.81}$Fe$_{0.19}$/Pb point contact. 
(b) The zero-bias conductance peak for a Ni$_{0.81}$Fe$_{0.19}$/Pb point contact. 
(c) The zero-bias conductance peak for a Ga$_{0.94}$Mn$_{0.06}$As/Pb point contact.
In (d), the previously measured PCAR spectrum with a broadened superconducting 
energy gap for a Fe/Pb point contact\cite{Strijkers}, together with the fit to the extended
BTK model\cite{Strijkers} (the dashed line), is also shown.}
\end{figure}

The physical origins of the dips, ZBCP, and large zero-bias enhancement
have been studied before. 
For example, Strijkers {\it et al.}\cite{Strijkers} investigated the dip structures
by introducing an additional proximity-induced superconducting energy gap.
On the other hand, intergrain Josephson effect\cite{Shan} and Maxwell resistance\cite{Sheet} were also regarded
as possible sources of both dips and large zero-bias enhancement.
Schmidt {\it et al.}\cite{Schmidt} proposed formation of superconductor-insulator-superconductor (SIS) junctions
that would result in the ZBCP.
Indeed, the geometry of a point contact may lead to the formation of a SIS junction.
However, the ZBCP also appears in planar superconductor-insulator-normal metal (SIN) junctions\cite{Mao}.
In ref. \cite{Geresdi}, the ZBCP was explained by multiple phase-coherent reflections
and proximity-induced Josephson effect. However, this explanation is inapplicable for the material with large exchange splitting.
Furthermore, the authors did not provide any model to fit their experimental data.
Interestingly, the Andreev bound states\cite{TanakaKashiwaya} have also been considered as a source of the ZPCP.
A recent theoretical work further showed that an Andreev bound state could be viewed as a topological edge state\cite{Tanaka}. 
However, an Andreev bound state stems from the unconventional pairing symmetry and thus would not occur in s-wave superconductors.
Therefore, the origins of the dip structures, zero-bias enhancement, and ZBCP are still not fully understood.

In addition to the above-mentioned anomalous features,  for some experimental data,
the value of the fitted model parameters for the extended BTK model of Strijkers {\it et al.}\cite{Strijkers},
e.g., $\Delta$, is much larger than the well-known value.
For example, in Fig. 1(d), the energy gap of Pb is 1.35 meV, whereas the fitted value is 1.93 meV.
Such PCAR spectra which exhibit a broadened energy gap, are called broadened spectra. 
However, two plausible mechanisms: thermal effects\cite{Kant,Auth,Bugoslavsky,Baltz}
and a spreading resistance\cite{Woods,Chiang, Chen2}, have been proposed to explain the broadened spectra.

In this paper, we investigate the transport mechanism of a ferromagnet-superconductor interface
and develop a modified BTK model to
better describe the physics at the interface and
account for anomalous PCAR spectra.
We solve a 3-D Bogolubov de-Gennes equations, assuming no scattering along the transverse direction.
We find that evanescent waves form for the large-angle incidence.
We point out that these evanescent waves might lead to charge inhomogeneity
and an electric dipole layer occurs at the interface.
Furthermore, we take the effect of charge inhomogeneity into account by modifying the barrier strength $Z$.
Interestingly, by doing that, we establish a new 3-D BTK model
which can describe the anomalous conductance spectra with dip structures,
a zero-bias conductance peak, and large zero-bias enhancement.
In other words, we show that
all anomalous conductance spectra have the same origin: transport induced charge inhomogeneity.
We also discuss the temperature-dependence of the dip structures.
Futhermore, we fit the experimental results using this new model including the effect of spreading resistance.
We find that not only the ZBCP spectra but also unsual spectra could be fitted well
by our model. 

\begin{figure}
\includegraphics[width=8cm]{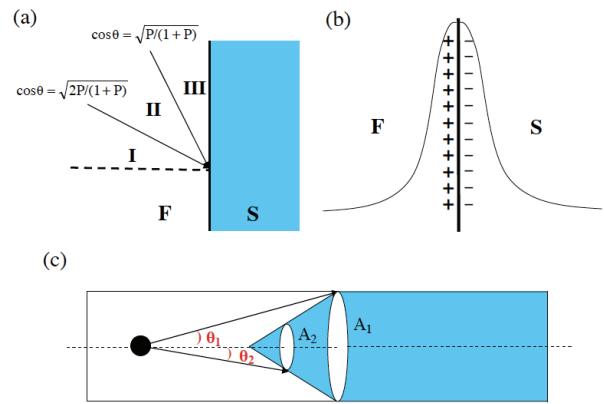}
\caption{(Color online) (a) For the spin-up incident electron, the incident region could be
divided into three parts. Here, $\theta$ is the incident angle.  
In region I, all wave functions are traveling waves.
In region II, the wave function of an Andreev reflected hole would decay.
In region III, all the wave functions of Andreev reflected hole, 
hole-like, and electron-like states are evanescent. 
The boundaries of the three regions depend on the spin polarization ($P$).
(b) A schematic plot of charge distribution near the interface due to the evanescent Andreev reflection and evanescent quasiparticle transmission.
(c) The relation between contact area and incident angle.
The white region on the left represents the ferromagnet while the gray (blue) region on the right indicates the superconducting tip.
The black circle denotes the incident electron.}
\end{figure} 

\section{Theoretical model}
We begin with the 3-D Bogolubov de-Gennes Hamiltonian.
\begin{align} 
&\left(\begin{array}[c]{cc}
  \hat{H_0}(\mathbf{r})-\rho_\sigma h(\mathbf{r}) & \Delta(\mathbf{r}) \\
   \Delta^*(\mathbf{r}) & -(\hat{H_0}+\rho_\sigma h({\mathbf r}))
  \end{array}\right)\left(\begin{array}[c]{cc}
   f(\mathbf{r})\\g(\mathbf{r})
  \end{array}\right)\notag\\
&=E\left(\begin{array}[c]{cc}
   f(\mathbf{r})\\g(\mathbf{r})
  \end{array}\right),\notag\\
&\hat{H_0}=-\frac{\hbar^2}{2m}\nabla^2+U\delta(x)-E_{F}.
\end{align}
Here, exchange potential $h(\mathbf{r})=E_{ex}\Theta(-x)$ and pair potential $\Delta(\mathbf{r})=\Delta e^{{\it i}\phi}\Theta(x)$,
the $x$-axis is normal to the interface.
The ferromagnet and superconductor are on the left and right side of the interface, respectively (see Fig. 2).
$\sigma$ denotes up (down) spin and $\rho_{\uparrow}$ ($\rho_{\downarrow}$) is 1 (-1).
Like other BTK formalisms, we model the interface scattering by a delta-function potential $U\delta(x)$.
In addition, we define the spin polarization
\begin{equation}
P=\frac{(N_{\uparrow}\upsilon_{F\uparrow}-N_{\downarrow}\upsilon_{F\downarrow})}
{(N_{\uparrow}\upsilon_{F\uparrow}+N_{\downarrow}\upsilon_{F\downarrow})},
\end{equation}
where $N_{\uparrow}$($N_{\downarrow}$) and $\upsilon_{F\uparrow}$($\upsilon_{F\downarrow}$)
are density of states and Fermi velocity for spin-up (down) electrons, respectively. We obtain $P=E_{ex}/E_{F}$.
Furthermore, we assume the $E$ and $\sqrt{E^2-\Delta^2}$ terms are negligible and hence obtain
the wave vectors for electrons, holes, and quasiparticles in superconductor $|k_{e\sigma}|\approx k_F\sqrt{1+\rho_{\sigma}P}$,
$|k_{h\sigma}|\approx k_F\sqrt{1-\rho_{\sigma}P}$, $|k_s|\approx k_F$, respectively.
Here, $k_F=\sqrt{2m(E_F+E_{ex})}/\hbar$.

\subsection{Incident angle and Andreev reflection}
We consider all possible incident angles which vary from $0$ to $\pi/2$\cite{Mortensen,Chalsani}
and assume no scattering parallel to the interface
(i.e. the transverse component of wave vector $k_{\\}$ is the same
in the ferromagnet and superconductor) \cite{Mortensen}.
We then derive the eigenstates of the Bogolubov de-Gennes Hamiltonian.

By matching the transverse components of wave vectors at the boundary, 
we find that with for the spin-up incident electrons, 
the forms of wave functions could be separated into three distinct regions, 
as shown in Fig. 2(a).
When the angle of incidence is less than $\cos^{-1}[\sqrt{2P/(1+P)}]$, wave functions do not decay.
Conversely, when the incident angle is larger than $\cos^{-1}[\sqrt{2P/(1+P)}]$, the Andreev reflected hole becomes evanescent.
Increasing the incident angle above $\cos^{-1}[\sqrt{P/(1+P)}]$,
the transmitted states in the superconductor (i.e. electron-like and hole-like states) start to decay and cannot propagate.
Therefore, we use the two boundary conditions: $\Psi_F(x=0^-)=\Psi_S(x=0^+)$ and $d\Psi_S/dx|_{x=0^-}-d\Psi_F/dx|_{x=0^+}=2Zk_F\Psi(0)$ 
to match the wave functions for the three regions, separately.
Here, $Z$ is the barrier strength.
Following the original BTK model\cite{Blonder}, $Z$ is defined by $mU/\hbar^2k_F$.

In region I (i.e. $\cos^2\theta>2P/(1+P)$), the wave functions in the ferromagnet and superconductor read as follows.
\begin{equation*}
\Psi_F=\left(\begin{array}[c]{cc}
   1\\0\end{array}\right)e^{{\it i}k_{ex}x}+a\left(\begin{array}[c]{cc}
   0\\1\end{array}\right)e^{{\it i}k_{hx}x}+b\left(\begin{array}[c]{cc}
   1\\0\end{array}\right)e^{-{\it i}k_{ex}x},
\end{equation*}
\begin{equation}
\Psi_S=c\left(\begin{array}[c]{cc}
   u_0\\v_0\end{array}\right)e^{{\it i}k_{sx}x}+d\left(\begin{array}[c]{cc}
   v_0\\u_0\end{array}\right)e^{-{\it i}k_{sx}x}.
\end{equation}

In region II (i.e. $P/(1+P)<\cos^2(\theta)<2P/(1+P)$),
unlike region I, we take into account the evanescent hole state and hence $\Psi_F$ is written as
\begin{equation}
\Psi_F=\left(\begin{array}[c]{cc}
   1\\0\end{array}\right)e^{{\it i}k_{ex}x}+a\left(\begin{array}[c]{cc}
   0\\1\end{array}\right)e^{\rho_{hx}x}+b\left(\begin{array}[c]{cc}
   1\\0\end{array}\right)e^{-{\it i}k_{ex}x}.
\end{equation}
However, $\Psi_S$ is the same as in region I.

In region III (i.e. $\cos^2\theta<P/(1+P)$), in addition to the evanescent Andreev reflected hole, the transmitted waves decay exponentially
and hence we have to rewrite $\Psi_S$ as,
\begin{equation}
\Psi_S=c\left(\begin{array}[c]{cc}
   u_0\\v_0\end{array}\right)e^{-\rho_{sx}x}+d\left(\begin{array}[c]{cc}
   v_0\\u_0\end{array}\right)e^{-\rho_{sx}x}
\end{equation}
where $u_0$ and $v_0$ have the same definition as in ref. \cite{Blonder}.
The wave vectors in the $x$-direction are $k_{ex}=k_{F}(1+P)^{1/2}\cos\theta $, $k_{hx}=k_{F}[(1+P)\cos^2\theta-2P]^{1/2}$,
and $k_{sx}=k_{F}[(1+P)\cos^2\theta-P]^{1/2}$.
The decay parameters $\rho_{hx}=k_{F}[2P-(1+P)\cos^2\theta]^{1/2}$,
and $\rho_{sx}=k_{F}[P-(1+P)\cos^2\theta]^{1/2}$.
$k_F$ denotes the Fermi wave vector.

As to the spin-down incident electron, the wave functions never decay regardless of how large the incident angle is.
Therefore, the forms of wave functions are the same as that in region I
except the wave vectors are different.
The wave vectors in the $x$-direction for the spin-down incident electron are
$k_{ex}=k_{F}[1-P]^{1/2}\cos\theta$, $k_{hx}=k_{F}[(1-P)\cos^2\theta+2P]^{1/2}$, and $k_{sx}=k_{F}[(1-P)\cos^2\theta+P]^{1/2}$.

The wave vectors of the incident electron and the Andreev reflected hole are not equal.
This indicates that the reflected hole does not retrace the path of the incident electron,
as in the usual Andreev reflection process.
As pointed out in ref. \cite{Kashiwaya}, for the 3-D model, in the presence of the exchange interaction,
the retro-reflectivity of Andreev reflection would be broken.
Recently, it was shown that specular Andreev reflection may happen
in graphene metal-superconductor structures\cite{Beenakker1,Hsu}.
As to the decaying waves at larger incident angle, they stand for the states localized near the interface\cite{Katsnelson}.
After matching wave functions at the interface, we can obtain the amplitudes of the
Andreev reflection $a$, normal reflection $b$, electron-like transmission $c$,
and hole-like transmission $d$ and hence the probabilities for each region.

In addition, it should be noted that evanescent Andreev reflection and evanescent quasiparticles transimission would
also happen for a normal metal-superconductor interface.
The evanescent Andreev reflection and quasiparticle transmission occur when $k_e\sin\theta>k_h$ and $k_e\sin\theta>k_s$.
When $k_e=k_h$ ($k_e=k_s$), $k_e\sin\theta\leq k_h$ [$k_e\sin\theta\leq k_s$] for any incident angle.
Therefore, the inequality of wave vector is a necessary condition for the formation of the evanescent states.
For higher spin polarization (ferromagnet), the inequality of wavectors is mainly caused by the exchange splitting.
Hence, it is reasonable to neglect the terms of $E$ and $\sqrt{E^2-\Delta^2}$ for mathematical simplication.
However, when the spin polarization is close to zero, $E$ and $\sqrt{E^2-\Delta^2}$ are not negligible compared to the exchange splitting.
For $P\approx0$, if we keep $E$ and $\sqrt{E^2-\Delta^2}$ in wave vectors,
the wavectors of the incident electron and Andreev reflected hole are given by
$(\hbar k_e)^2/2m = (\hbar k_F)^2/2m +E$, $(\hbar k_h)^2/2m =(\hbar k_F)^2/2m -E$, $(\hbar k_s^+)^2/2m =(\hbar k_F)^2/2m +\sqrt{E^2-\Delta^2}$, and $(\hbar k_s^-)^2/2m =(\hbar k_F)^2/2m -\sqrt{E^2-\Delta^2}$.
Here, $k_s^+$ and $k_s^-$ are the wave vectors of electron-like and hole-like states, respectively.
In the presence of $E$ and $\sqrt{E^2-\Delta^2}$, therefore, the evanescent Andreev reflection and quasiparticle
transmission would appear in the normal metal-superconductor point contacts as well.
Our mathematical formulae presented in this paper are applicable for the ferromagnet-superconductor interfaces
only. In the future, we will study the evanescent transport at normal metal-superconductor interfaces.

\subsection{Modification of barrier strength}


\begin{table*}
\caption{The analytical forms of $(a^*a)_{II}$, where $a$ is the complex coefficient in Eq. (4), and $(c^*c-d^*d)_{III}(u^*_0u_0-v^*_0v_0)$ in Eq. (5).
Here, $u_0^2=1-v_0^2=\frac{1}{2}(1+\sqrt{1-\Delta^2/E^2})$ and $\lambda=P/(1+P)\cos^2\theta$.}
\begin{tabular}{lll} 
\hline
           & $(a^*a)_{II}$ & $(c^*c-d^*d)_{III}(u^2_0-v^2_0)$  \\ \hline \\
$|E|<\Delta$ & $\frac{4(1-\lambda)\Delta^2}{\left[\left(\sqrt{2\lambda-1}+2Z_1\sqrt{\lambda/P}\right)\sqrt{\Delta^2-E^2}\right]^2
+\left[\left(1-\lambda+2Z_1\sqrt{(2\lambda-1)\lambda/P}+4Z^2_1\lambda/P\right)\sqrt{\Delta^2-E^2}
-\sqrt{3\lambda-2\lambda^2-1}|E|\right]^2}$ & 0  \\ \\
$|E|>\Delta$ & $\frac{16(1-\lambda)u^2_0v^2_0}{\left[\left(1-\lambda+2Z_1\sqrt{(2\lambda-1)\lambda/P}
+4Z^2_1\lambda/P\right)(u^2_0-v^2_0)+\sqrt{1-\lambda}\right]^2
+\left[\left(\sqrt{2\lambda-1}+2Z^2_1\sqrt{\lambda/P}\right)(u^2_0-v^2_0)+\sqrt{3\lambda-2\lambda^2-1}\right]^2}$
& $\frac{4}{\left(1-\lambda+2Z^2_1\sqrt{\lambda/P}\right)^2+1}$  \\ \\
\hline
\end{tabular}
\label{tableD}
\end{table*}

In the original BTK theory\cite{Blonder}, it was assumed that the applied bias voltage has no effect on the barrier potential or $Z$.
However, this assumption was not based on any physical argument and may be invalid.
From the above calculation, we find that evanescent waves form for the large-angle incidence.
Therefore, the likely influence of evanescent states on the barrier potential of the interface should be considered.
The evanescent wave represents the spatially inhomogeneous distribution of particles.
Hole and quasiparticles carry charges.
This implies that charge may not spread uniformly and could be localized near the interface instead.
This charge inhomogeneity would affect the transport of other incident electrons via electrostatic interaction.

In the evanescent Andreev reflection process, the incident electron near the interface would take another electron away from the ferromagnet
to the superconductor, and an evanescent hole is therefore left in the ferromagnet.
The movement of a hole in one direction implies the movement of an electron in the opposite direction.
The evanescent Andreev reflected hole does not carry any current.
That is, the removed electron does not travel either after it is transferred into the superconductor.
Therefore, via the evanescent Andreev reflection, positive charges would prefer to reside on the ferromagnetic side of interface.
Conversely, negative charges would be localized on the superconducting side of the interface.
Hence, an electric dipole layer forms, as shown in Fig. 2(b).
Similarly, electric dipole layer could also occur via the evanescent quasiparticle transmission
because the charges of incident electrons are transferred into the superconductor via the formation of decaying quasiparticles
and localized positive charges appear in the ferromagnet near the interface.
The related study on dipole layer at metal-superconductor interface was conducted in ref. \cite{Nikolic}.
In addition, as shown in Fig. 2(a), evanescent transport happens at larger incident angles.
Furthermore, in Fig. 2(c), an electron at the black circle may be incident
at the ferromagnet-superconductor interface along various directions.
However, the maximum possible incident angle is limited by contact area.
For example, the contact area $A_1$ is larger than $A_2$
and hence the maximum possible incident angle $\theta_1>\theta_2$.
The maximum possible incident angle increases with contact area.
Therefore, we can infer that the effect of electric dipole layer would be more striking when the contact area is larger.

The electrostatic potential due to an electric dipole layer may be written in the form\cite{Jackson}
\begin{equation}
\Phi(\mathbf{x})=\frac{1}{4\pi\epsilon}\int_S D(\mathbf{x'})\mathbf{n}
\cdot\nabla'\left(\frac{1}{|\mathbf{x}-\mathbf{x'}|}\right)dS'.
\end{equation}
Here, $D(\mathbf{x'})$ is the surface dipole moment density and $\mathbf {n}$ is the direction of dipole moment of the layer.
Obviously, the electrostatic potential decreases with the distance between the observer and dipole layer,
and becomes quite strong when the observer is very close to the dipole layer.
The electric dipole layer due to charge transfer at the ferromagnet-superconductor interface should display the similar behavior
and hence has a maximum in the vicinity of the interface.
Furthermore, taking into account screening effect, the range of the potential could be finite.
In metals, the screening length is about a few Angstroms.
Therefore, it would be reasonable to model the screened short-range electrostatic potential as a $\delta$-function.
Using the definition of the charge density $Q=e(|f|^2-|g|^2)$ in ref. \cite{Blonder},
we find that the charge density of an evanescent hole is proportional to $a^*a$.
Therefore, the barrier potential due to the contribution of an evanescent Andreev reflection is linear in $a^*a$.
However, the strength of barrier potential is difficult to determine,
because the $\delta$-function is unphysical in reality\cite{Kant}
and merely for the purpose of mathematical simplification.
Therefore, we set the strength of barrier potential equal to $U_2(a^*a)$,
where $U_2$ is an undetermined constant.
We use the similar method to model the barrier potential
due to evanescent quasiparticles transmission.
Furthermore, using Eq. (6), we can find that the electrostatic potential energy
for incident electrons due to these evanescent states is negative.
Because $a$ in the region III is found to be $0$,
the barrier potential is modified as $\tilde{U}=U_1-U_2(a^*a)_{II}-U_3(c^*c-d^*d)_{III}(u^*_0u_0-v^*_0v_0)$.

We can derive the modified barrier strength
\begin{equation}
\tilde{Z}=Z_1-Z_2(a^*a)_{II}-Z_3(c^*c-d^*d)_{III}(u^*_0u_0-v^*_0v_0).
\end{equation}
Here, $Z_1$ is related to the scattering due to the contact potential, impurity,
and tunneling barrier at the interface, as the previous BTK defined.
$Z_2$ is the barrier strength that the evanescent Andreev reflection contributes.
$Z_3$ is caused by the Coulomb potential due to the evanescent quasiparticles transmission.
We assume that the localized charges affect only the transport of
the incident spin-up electrons in region I and the incident spin-down electrons.
In other words, we neglect the influence the localized states in regions II and III on the transport of each other.
The barrier potential for the spin-up incident electrons in regions II and III is not altered by the charge inhomogeneity.
Therefore, we use this modified $\tilde{Z}$ for the incident spin-up electrons in region I
and the incident spin-down electrons.
For the spin-up incident electrons in regions II and III, the barrier strength is equal to $Z_1$.
The $(a^*a)_{II}$ and $(c^*c-d^*d)_{III}(u^*_0u_0-v^*_0v_0)$ are shown in Table. I.

\subsection{Calculation of conductance}
Finally, by averaging over incident angles between 0 and $\pi/2$, as in ref. \cite{Chalsani},
we obtain the average Andreev reflection $\bar{A}_\uparrow$($\bar{A}_\downarrow$) and
normal reflection $\bar{B}_\uparrow$($\bar{B}_\downarrow$) probabilities for the spin-up (down) electron.
Using these probability coefficients, we can calculate the conductance of a point contact as
in the BTK model\cite{Blonder}.
The conductance is given by
\begin{align}
G(V) &=N_{\uparrow}e^2\mathcal{A}\upsilon_{F\uparrow}\int^{\infty}_{-\infty}-\frac{\partial f(E-eV)}{\partial E}(1+\bar{A_\uparrow}-\bar{B_\uparrow})dE\notag\\
&+N_{\downarrow}e^2\mathcal{A}\upsilon_{F\downarrow}\int^{\infty}_{-\infty}-\frac{\partial f(E-eV)}{\partial E}(1+\bar{A_\downarrow}-\bar{B_\downarrow})dE
\end{align}
Here, both $N_\uparrow/N_\downarrow$ and $\upsilon_{F\uparrow}/\upsilon_{F\downarrow}$ are equal to $\sqrt{(1+P)/(1-P)}$.
Furthermore, we normalize the conductance by $G_{N}$, which is equal to $G(V)|_{eV\gg\Delta}$.

In order to explain the spectra broadening effect, we also take into account the extra resistance factor $\Gamma$,
which results from the spreading resistance.
In the presence of spreading resistance, the normalized conductance $G_{FS}/G_{FN}$ is modified as
\begin{align}
\left(\widetilde{\frac{G_{FS}}{G_{FN}}}\right) &=\frac{1+\Gamma}{(G_{FS}/G_{FN})^{-1}+\Gamma},\notag\\
\mbox{with } \Gamma &= R_{s}/R_{FN},
\end{align}
where $R_{FN}=1/G_{FN}$ represents the resistance at large bias voltage, while $R_{s}$ denotes the spreading resistance.
We use this formalism to fit the experimental data and the results are presented in Sec. IV.

\section{Experiments}

We fabricated some permalloy ($\mbox{Ni}_{0.81}\mbox{Fe}_{0.19}$)
thin films on Si substrates by DC magnetron sputter system.
The thickness of samples are about $100$ nm.
The bulk resistivity is $\sim$287 (n${\Omega}$m).
To perform the PCAR measurements, a differential screw was used to bring a fine tip onto the sample surface.
Fine Pb tips fabricated by mechanical grinding were pressed onto the sample repeatedly to ensure
good contact between the tips and the samples.
To minimize thermal fluctuation, conductance-voltage curves were taken by the differential technique\cite{Keithley}
in a helium cryostat for 4.2 K measurements.
The conductance curves were normalized by the high bias values.
Each sample was measured 50 times on different contact condition.

The contact size was estimated from the quasiclassical formula of contact resistance,
applicable to both ballistic and diffusive transport\cite{Geresdi,Branislav}
\begin{align}
R&=(1+Z^2)\left[\frac{4\rho\ell_{m}}{3{\pi}d^2}+\gamma\left(\frac{\ell_{m}}{d}\right)\frac{\rho}{2d}\right]
\end{align}
where $d$ is the diameter of the contact, $\rho$ is the resistivity of the material,
$\gamma(\ell_{m}/d)$ is a factor in the order of unity.
From this contact resistance formula we can calculate the contact diameter in the range of about $2\sim19$ nm.
Selected PCAR spectra with dip structures and ZBCP from our permalloy thin films are displayed in Fig. 1(a) and (b),
respectively.
Some of the previously measured dilute magnetic semiconductor samples\cite{Chiang} (Fig.1 (c)),
which have much larger resistivities, can also be analyzed with this model.

\begin{figure}
\includegraphics[width=6.0cm]{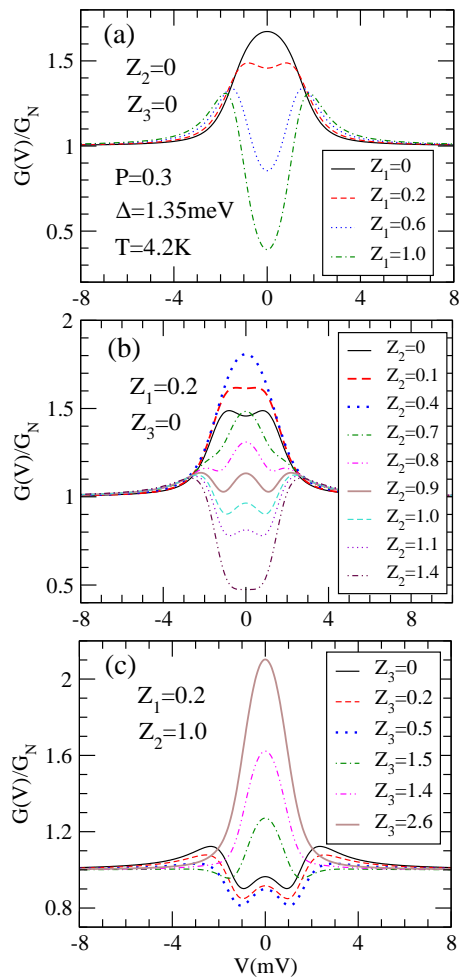}
\caption{(Color online) (a) The conductance spectra for different origin barrier strength
at fixed spin polarization $P=0.3$ when $Z_2$ and $Z_3$ are not applied.
(b) The effect of evanescent-hole-induced barrier strength $Z_2$ on conductance spectra.
(c) The effect of evanescent-quasiparticle-induced barrier strength $Z_3$ on conductance spectra.
The parameters $P$, $\Delta$, and $T$ in (b) and (c) are the same as (a).}
\end{figure}

\section{Results and discussion}

\subsection{Effects of barrier strengths}
In Fig. 3, the calculated conductance spectra under three different barrier strengths are displayed in order to
examine the influence of the barrier strength.
We find from Fig. 3(a) that when the effect of charge inhomogeneity are not taken into account (i.e., $Z_2=0$ and $Z_3=0$),
the normalized conductance at low bias decreases with increasing barrier strength $Z_1$
and two peaks appear symmetrically about zero bias, like the previous BTK models\cite{Blonder,Strijkers}.
This was already attributed to the suppression of Andreev reflection by the barrier strength\cite{Blonder}.
In Fig. 3(b), it is shown how the evanescent Andreev reflection [through barrier strength (i.e. $Z_2\neq0$)]
affect the conductance spectra. 
Clearly, the low-bias conductance would increase initially, with increasing $Z_2$, as shown by the curves with $Z_2=0.1$ and $Z_2=0.4$,
but, however, would decrease upon further increasing $Z_2$.
From Eq. (7), we can infer that $Z_2$ would reduce the effective barrier strength $\tilde{Z}$ and hence increases the low-bias conductance.
However, if $Z_2$ is large, the effective barrier strength becomes negative.
The negative potential barrier implies a potential well rather than a potential barrier. 
When the effective barrier strength becomes negative, the depth of the well increases with increasing $Z_2$
and hence the conductance would be reduced by $Z_2$.
Moreover, when the conductance increases with $Z_2$ (i.e. the curves for $Z_2=0,0.1,$ and $0.4$),
the dip at the zero-bias would diminish gradually and eventually a single peak occurs at zero bias (see the curve for $Z_2=0.4$).
The shape of conductance spectra for $Z_2=0.4$ is similar to that of the curve in the clean regime
(i.e. the curve for $Z_1=0, Z_2=0, Z_3=0$, shown in Fig. 3(a)).
However, in comparison with the curves with no barrier, the height of the peak for $Z_2=0.4$ is higher.
The heights of peaks for $Z_2=0.4$ and for no barrier are 1.81 and 1.67, respectively.
In addition, we find that the peak for $Z_2=0.4$ is sharper.
We also notice that $Z_2$ can generate double dips, like the curves for $Z_2=0.8,0.9,1.0$ and $1.1$.

In Fig. 3(c), the effect of the evanescent-quasiparticle-induced barrier strength $Z_3$ is shown.
In contrast with $Z_2$, $Z_3$ would reduce the low-bias conductance first and then raise it.
It is clear from Table I that the coefficient $(c^*c-d^*d)_{III}(u_0^*u_0-v_0^*v_0)\neq0$ for $|E|>\Delta$
and the $(a^*a)_{II}$ approaches zero at high energy.
Therefore, $G_N$ is insensitive to $Z_2$ but may be influenced by $Z_3$.
Similar to $Z_2$, for smaller $Z_3$, $Z_3$ reduces the height of the barrier at high bias
and hence $G_N$ increases and the low-bias normalized conductance decreases.
For sufficiently large $Z_3$, a barrier is transformed into a well and $G_N$ decreases with $Z_3$.
Hence, the normalized conductance could be enhanced by increasing $Z_3$.
Furthermore, $Z_3$ could raise zero-bias conductance above 2 (see the curve for $Z_3=2.6$).

\begin{figure}
\includegraphics[width=8cm]{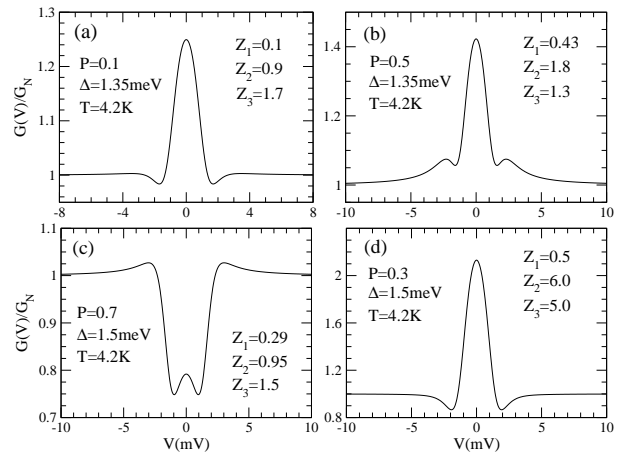}
\caption{The simulated anomalous conductance spectra with (a) dip structures,
(b), (c) zero-bias conductance peak, and (d) both dip structures and large zero-bias enhancement.}
\end{figure}

It is worth mentioning that, under the effect of charge inhomogeneity,
the zero-bias normalized conductance does not reach its maximum in the clean limit.
Indeed, the transmission of electrons into the superconductor should be obstructed by a barrier (or a well)
especially upon Andreev reflection, as shown in Fig. 5 of ref. \cite{Blonder}.
Furthermore, at high bias, electrons propagate into the superconductor mainly through electron- and hole- like transmission
and hence $G_N$ is independent of Andreev reflection probability.
Therefore, the zero-bias conductance $G(0)/G_{N}$ must decrease if $Z$ is a constant.
When $Z_2$ and $Z_3$ are included, the total barrier strength varies with voltage.
Electrons may feel a higher barrier or a deeper well at high bias
and therefore normalized conductance could increase in the presence of a barrier or a well.

\begin{figure*}
\includegraphics[width=14cm]{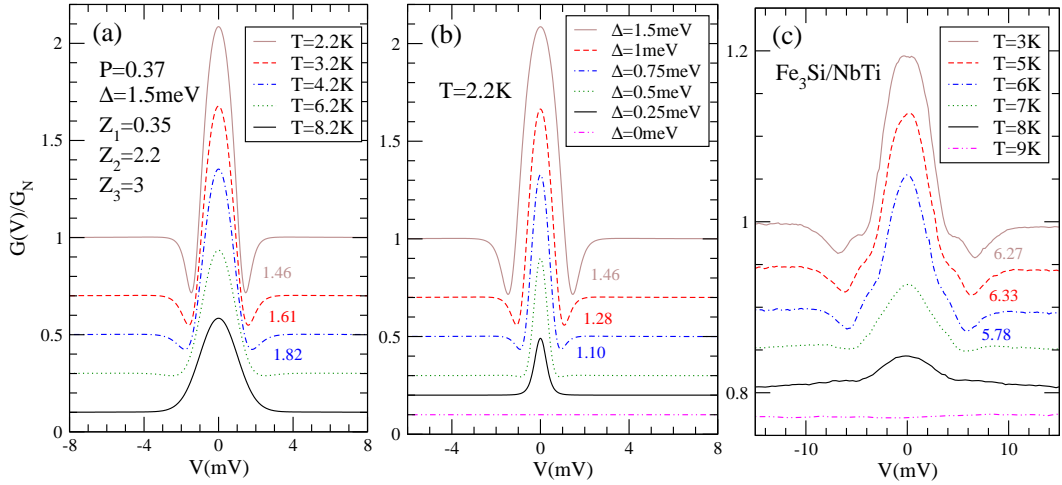}
\caption{(Color online) The temperature-dependence of dip structures.
(a) The calculated conductance spectra for different temperatures $T$ at fixed superconducting energy gap $\Delta$.
(b) The calculated conductance spectra for different superconducting energy gaps $\Delta$ at fixed temperature $T$.
(c) The measured point-contact Andreev reflection spectra for a Fe/Nb point contact at different temperatures 
from ref. \cite{Sheet}. 
All curves in (a)-(c) except the top curve are shifted downward for clarity.
The parameters $P$, $Z_1$, $Z_2$, and $Z_3$ in (b) are the same as (a).  
The numbers marked in (a), (b), and (c) are the positions of dips. }
\end{figure*}

\subsection{Anomalous spectra}
From the preceding section, we may conclude that
the evanescent-hole-induced barrier strength $Z_2$ could lead to zero-bias enhancement,
sharp central peak, occurrence of dips while the evanescent-quasiparticle-induced barrier strength $Z_3$
could substantially enhance the zero-bias conductance.
By applying $Z_2$ and $Z_3$, we can create the anomalous PCAR spectra for different spin polarizations,
which are very similar to experimental observations, as shown in Fig. 4.
In other words, all the observed anomalous characteristics can be reproduced in our model
with suitable values of the various barrier strength parameters.

Fig. 4(a) shows the anomalous spectra with the dip structures.
In experiments, the dips are usually located at $eV\gtrsim\Delta$\cite{Strijkers,Sheet}.
The dips occur at $V=\pm1.69$ mV in the present case.
In the presence of the spreading resistance,
the dips would move farther away from the central peak.
This is consistent with the experimental observations.
Fig. 4(b) and (c) show two different types of the ZBCP spectra. 
In Fig. 4(b), a higher conductance peak is located at zero bias
and two small peaks appear symmetrically on both side of it,
similar to the experimental results in ref. \cite{Geresdi}.
In Fig. 4(c), a zero-bias shallow hump appears at tunneling-like Andreev reflection spectra,
similar to the experimental results in refs. \cite{Srikanth,Kastalsky,Magnee,Quirion}.
The two ZBCP characteristics can both be obtained by adjusting $Z_2$,
as shown in the curves for $Z_2=0.8$ and $Z_2=0.9$ of Fig. 3(b).
Our results indicate that both dip structures and a ZBCP are due to the evanescent Andreev reflection.
In ref. \cite{Srikanth}, it was showed that the dip structures in the Andreev spectra [like Fig. 4(a)]
could be transformed into the ZBCP [like Fig. 4(b)] by varying contact area,
which is consistent with our results.
Adjusting contact area could alter the properties of interface, $i.e.$, $Z_1$, $Z_2$, and $Z_3$ and
hence different types of the PCAR spectra would occur.

Fig. 4(d) shows both the dip structures and large zero-bias enhancement.
In experiments, the dip structures are often accompanied by the large zero-bias enhancement\cite{Shan,Sheet}.
Our model can reproduce both the dip structures and large zero-bias enhancement.
In addition, we find from Fig. 3(b) that the dips appear when $Z_2$ reduces the zero-bias conductance.
This indicates that the unusual zero-bias enhancement results from $Z_3$.
In experiments, the dip structures, a ZBCP, and large zero-bias enhancement usually happen at large contact area (i.e. low resistance).
Charge inhomgeneity is also pronounced for large contact area, as seen in Fig. 3(c).
This supports the validity of our model.
However, it should be noted that the increase of conductance with contact area is due to
the increase of electron-transport channels instead of $Z_2$ or $Z_3$.
In the preceding subsection, it was illustrated how $Z_3$ reduces $G_N$ and $G(0)$
under the assumption of constant electron-transport channels.
In principle, the charge inhomogeneity could also happen for the planar geometry.
Therefore, our model could also be extended to the planar junctions.

\subsection{Temperature dependence of dip structures}
Let us now examine how the dip structures vary with temperature.
The superconducting energy gap decreases with increasing temperature\cite{Tinkham}.
We plot the normalized conductance for different values
of temperature and energy gap in Fig. 5(a) and Fig. 5(b), respectively.
Fig. 5(a) indicates that the dips would be reduced gradually and also move away from the central peak as $T$ increases.
Fig. 5(b) shows that, with decreasing $\Delta$, the dips would diminish gradually, too, but move towards the central peak.
We can infer that the dips would disappear eventually when the temperature is raised.
Moreover, from Fig. 2-2 in ref. \cite{Tinkham}, near $T=0$, the energy gap is almost insensitive to $T$.
If $T$ is larger, $\Delta$ would decrease with $T$ more rapidly.
When $T$ is increased up to the critical temperature $T_c$, $\Delta$ drops to zero.
In other words, we predict that, with the temperature rising,
the position of a dip should increase if temperature is low enough.
Conversely, at higher temperatures, due to the rapid decrease of $\Delta$,
the position of a dip could decrease with increasing temperature.
Fig. 5(c) shows the measured temperature dependence of the dips for a Fe/Nb point contact\cite{Sheet},
which should correspond to the theoretical results in Fig. 5(b).
Many other experimental works have shown the same temperature dependence of dip structures for a wide variety
of combinations between metals and superconductors\cite{Nowack,Mao,Shan,Gourgy,Daghero}.
Furthermore, it was pointed out in the preceding subsection that dip structures and the ZBCP have the same physical origin.
Therefore, indeed, the two local minima in the ZBCP spectra would be equivalent to the dip structures.
Kastalsky {\it et al.} \cite{Kastalsky} observed ZBCP spectra in the temperature range of $T=0.5-1.7$K. 
Their data showed the two local minima in the conductance spectra moved to higher bias as temperature increased.
Therefore, we infer that the dips may also move away from the central peak, as shown in Fig. 5(a).

\begin{figure}
\includegraphics[width=7cm]{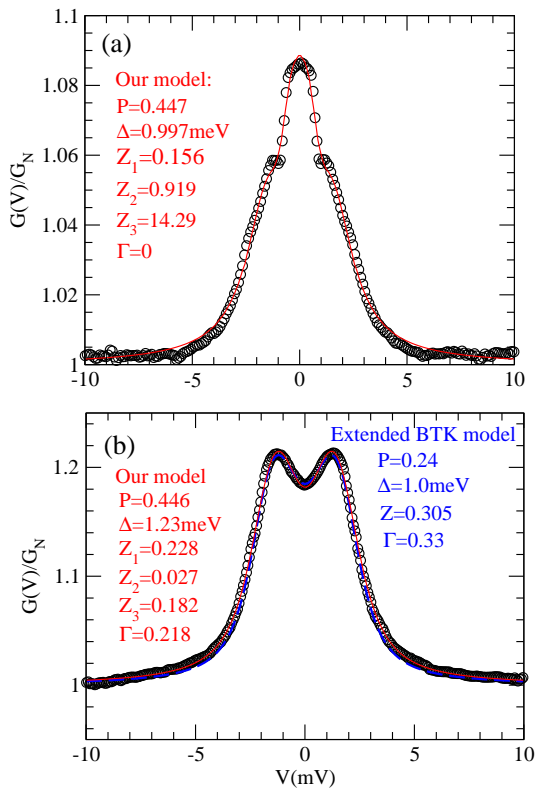}
\caption{(Color online) The conductance spectra for Ni$_{0.81}$Fe$_{0.19}$/Pb
point contacts measured at 4.2K (black circles) 
with contact resistance of (a) $R=3.80$ $\Omega$ and (b) $R=8.83$ $\Omega$.
The solid (red) and dashed (blue) lines are the fits using our model (see Sec. II)
and the extended BTK model\cite{Strijkers}, respectively. 
The fit value of temperature is fixed at 4.2 K.} 
\end{figure}

\subsection{Fitting experimental spectra}

We performed the least squared fit to selected
experimental data using our model and compared with the extended BTK model.\cite{Strijkers}
In addition to the $P$, $Z_1$, $Z_2$, and $Z_3$, which are necessary in our model, $\Delta$, $T$,
and the extra resistance factor $\Gamma$ in Eq. (9) can also be used as fitting parameters to the experimental spectra.
To minimize the fitting parameters, we chose to fit the measurements obtained from samples immersed in
liquid Helium and fixed the temperature $T=4.2$ K.
At first, superconducting energy gap $\Delta$ for Pb was fixed at 1.23 meV and the $\Gamma$ was fixed at zero.
If the fitting result was not satisfactory, we would turn on $\Delta$ and/or $\Gamma$ as fitting parameters.

We find that, for small contact resistance ($R=3.80$ $\Omega$),
a zero-bias conductance peak could appear in the experimental curve [Fig. 6(a)].
However, for large contact resistance ($R=8.83$ $\Omega$), we do not observe the ZBCP
and an ordinary PCAR spectrum occurs, as shown in Fig. 6(b).
We also find from Fig. 6(a) that, in contrast to the extended BTK model\cite{Strijkers}
which could not reproduce the shoulder structures,
our model could fit the ZBCP spectrum very well.
Fig. 6(b) shows that our model also fits usual spectra well.
Although the usual spectra could be fitted by the extended BTK model\cite{Strijkers} well,
the resultant parameter $P=0.24$ is significantly
smaller than that from the previous measurements\cite{Soulen,Nadgorny,Paraskevopoulos,Veerdonk}.
The spin polarization of permalloy has been determined by both PCAR\cite{Soulen,Nadgorny} and
spin-polarized tunneling\cite{Paraskevopoulos,Veerdonk} experiments, and was found to be in the range of 0.3-0.5.
In our model, the fitted value of $P\simeq0.45$ agrees well with these previous measurements.
This indicates that, in our measurements, charge inhomogeneity may happen and,
therefore, the extended BTK model could not explain the experimental data in Fig. 6(b).

\section{Conclusions}
In conclusion, we have developed a modified 3-D BTK model
by taking into account the effect of transport induced charge inhomogeneity on the barrier potential
at ferromagnet-superconductor interfaces.
To this end, we add two new parameters, namely, the evanescent hole-induced barrier strength $Z_2$ and
evanescent quasiparticle-induced barrier strength $Z_3$
to describe the modified barrier strength.
We find that $Z_2$ would give rise to the dip structures and zero-bias conductance peak,
while, on the other hand, $Z_3$ would enhance the zero-bias conductance by decreasing the normal-state conductance $G_N$.
We have, therefore, offered a possible explanation for the anomalous conductance spectra which have been long standing problems
in the field of PCAR spectroscopy.
Furthermore, the results of our model fits to the experimental data have a good
agreement with the measured ZBCP spectra as well as normal conductance spectra. 
Nevertheless, the standard $\delta$-function form adopted here for the barrier potential at point contacts could be an oversimplication.
Further theoretical and experimental works are certainly needed to fully understand these important issues
in the PCAR spectroscopy.

\section*{Acknowledgments}
The authors thank Hao-Chun Lee for valuable discussion.
The authors also acknowledge financial supports from the National Science Council, Academia Sinica, and National Center for Theoretical Sciences of Taiwan as well as Center for Quantum Science and Engineering, National Taiwan University (CQSE-10R1004021).


\end{document}